\documentstyle[aps,preprint]{revtex}
\begin{document}
\tightenlines
\draft
\title{Coarse-grained field wave function in stochastic inflation} 
\author{Mauricio Bellini\footnote{E-mail address: mbellini@mdp.edu.ar}}
\address{Instituto de F\'{\i}sica y Matem\'aticas,
Universidad Michoacana de San Nicol\'as de Hidalgo, \\
AP: 2-82, (58041) Morelia, Michoac\'an, M\'exico}
\maketitle
\begin{abstract}
The wave function
for the matter field fluctuations in the infrared sector
is studied within the framework of inflationary cosmology. 
These fluctuations
are described by a coarse-grained field which takes into account only the
modes with wavelength much bigger than the size of the Hubble horizon. 
The case of a power-law expanding universe is considered and it is found
that the relevant phase-space 
($\phi_{cg},P_{\phi_{cg}}$) remains coherent under certain circumstances. 
In this case
the classical stochastic treatment for matter field fluctuations is not
valid, however, for $p>4.6$, the system loses its coherence and
a classical stochastic approximation is allowed.
\end{abstract}
\vskip 2cm

\section{Introduction}

A standard mechanism for galaxy formation is the amplification of primordial
fluctuations by the evolutionary dynamics of spacetime.
The inflationary cosmology is based on the dynamics of a quantum field
undergoing a phase transition\cite{1}. The exponential expansion of the scale
parameter gives a scale-invariant spectrum naturally. This is one of the many 
attractive features of the inflationary universe, particulary with regard
to the galaxy formation problem\cite{2} and it
arises from the fluctuations of the
inflaton, the quantum field which induces inflation.
This field can be semiclassically expanded in terms of its
expectation value plus other field, which describes the quantum 
fluctuations\cite{3}.
The quantum to classical transition of quantum fluctuations has been
studied in thoroughly\cite{4}. 

The infrared matter field fluctuations are
classical and can be described by a coarse-grained field which takes
into account only wavelengths larger than the Hubble radius. The dynamics
of this coarse-grained field is described by a second order stochastic
equation, which can be treated using the Fokker-Planck 
formalism.
This issue has been the subject of intense work during the last two decades
\cite{Habib,BCMS,Mijic,CMS}.
A different approach is the quantum mechanical treatment of the coarse-grained
field\cite{HM}, where
the fluctuations are described by means
of a time dependent quadratic potential with a linear external stochastic
force. Even though an isolated system described by the 
Schr\"odinger equation cannot
lose its coherence, the
coase-grained field may evolve from a pure to a mixed state. One way
to realize coarse-graining is to let the system interact with an
environment. This consists of all the fields whose evolution we
are not interested in. The state of the system is obtained by tracing over 
all possible states of the environment. Even if the state
describing the system plus environment is pure, the
state of the system alone will in general be mixed. 
This is the case of the matter field fluctuations in inflationary
cosmology, where the system is given by the super Hubble modes
(infrared sector)
while the environment is given by the short-wavelength modes
(ultraviolet sector). For a supercooled expansion of the universe
the environment cannot be considered as a true environment because
it is not thermalized. A true environment appears in warm and
fresh inflationary scenarios\cite{ufa1,iuju} where
the inflaton field interacts with other particles of a thermal bath.
In these scenarios the environment is represented by the thermal bath
and the particles in it.

In this work, we aim at studying the phase-space decoherence of
the wave function that describes super
Hubble matter field fluctuations during inflation. Decoherence of the
phase-space is different to decoherence of the coarse-grained field.
While in the latter coarse-grained field decoherence consists in the
interchange of degrees of freedom between the infrared and ultraviolet sectors,
in the former there is interference between
the fluctuations of variables that describe the phase-space of
the quantum state. This issue has been treated before using either
the
Wigner function\cite{Po} or the Schr\"odinger formalism. 
It is well known that the evolution
of the redefined coase-grained field is described by a 
second-order stochastic
equation. 
This is considered in \cite{BCMS} for supercooled
inflation and also in \cite{Bell} for warm inflation.
The effective Hamiltonian related to this stochastic
equation can be expressed in such away that
the Schr\"odinger equation for the system can be written.
The wave function that describes this system is $\Psi(\chi_{cg},t)$, where
$\chi_{cg}$ denotes the coordinate and the variable
$P_{cg} \equiv \dot\chi_{cg}$ characterizes the
momentum of the phase-space. The interference 
between the the squared fluctuations
$\left<\chi^2_{cg}\right>$ and $\left<P^2_{cg}\right>$, which
arises from the coupling between the variables $\chi_{cg}$ and
$P_{cg}$ is the main feature to be analyzed in this paper.
It has been shown that in a de Sitter expansion of the universe
the phase-space of the quantum state remains pure\cite{HM}.
In this work the decoherence in the phase-space
for a power-law expanding universe in a globally Friedmann-Robertson-Walker
(FRW) metric is studied.

The paper is organized as follows: in section II, a 
revision of the inflationary formalism is done. In section III,
it is introduced the general quantum mechanical formalism for the 
coarse-grained field, which describe the redefined matter field
fluctuations on super Hubble scales.
In section IV, the wave function for the particular case of a
power-law expansion of the universe is studied. 
Finally, in section V, some
final remarks are given.

\section{Review of the inflationary formalism}

In a previous work\cite{BCMS} we justify the classical behaviour of the
order parameter on the basis of a semiclassical approach. The inflaton field
Lagrangian is: 
\begin{equation}
{\cal L}(\varphi ,\varphi _{,\mu })=-\sqrt{-g}\left[ \frac 12\left( g^{\mu
\nu }\varphi _{,\mu }\varphi _{,\nu }\right) +V(\varphi )\right] =a^3\left( 
\frac 12\dot{\varphi}^2-\frac 1{2a^2}(\nabla \varphi )^2-V(\varphi )\right),
\end{equation}
for a globally flat FRW metric, $ds^2=-dt^2+a(t)^2\ d\vec{r}^2$.
From here it follows the equation for the scalar field operator 
\begin{equation}
\ddot\varphi-\frac{1}{a^2}\nabla^2\varphi+3H\dot\varphi+V^{\prime}(%
\varphi)=0,
\end{equation}
and the Friedmann equation, written in terms of $H={\frac{\dot a }{a}}$, is 
\begin{equation}
H^2=\frac{4\pi}{3M_p^2}\left< \dot\varphi^2 +\frac{1}{a^2}
(\vec\nabla\varphi)^2+2V(\varphi) \right>,
\end{equation}
where the overdot represents the time derivative and $V^{\prime}(\varphi)={%
\frac{dV }{d\varphi}}$. We decompose the scalar field as its mean value plus
the quantum spatially inhomogeneous
fluctuations, $\varphi(\vec x,t) =\phi _{cl}(t)+\phi(\vec x,t) $
with $<\phi >=0$, up
to linear terms in $\phi $. Hence, the equations of motion reduce to
a set of two classical equations which give the evolution of the field 
$\phi_{cl}$ and the Hubble parameter. For simplicity, we
consider the Hubble parameter as classical: $H\equiv H(\phi_c) = {\dot a\over
a}$.
To be consistent with the 
FRW metrics, it is
assumed that $\phi _{cl}$ is a homogeneous field, and thus we have: 
\begin{eqnarray}
&&\ddot{\phi}_{cl}+3H\dot{\phi}_{cl}+V^{\prime }(\phi _{cl})=0,  \label{cl1}
\\
&&H^2=\frac{8\pi ^2}{3M_p^2}\rho,  \label{cl2}
\end{eqnarray}
where $V'(\phi_{cl}) = {dV(\phi_{cl}) \over d\phi_{cl}}$ and
$\rho =\frac 12\dot{\phi}_{cl}^2+V(\phi_{cl})$ is the
vacuum energy density.
The equation for the quantum fluctuations is
\begin{equation}\label{r6}
\ddot{\phi}-\frac 1{a^2}\nabla ^2\phi +3H\dot{\phi}+V''(\phi _{cl})\phi =0.
\label{ecop}
\end{equation}
In this last equation $H(\phi_{cl})$ and $V''(\phi_{cl})$
are given by equations (\ref{cl1}) and (\ref{cl2}), and they both 
are functions of $t$.

The characteristic timescale for the inflaton field can be defined by $\tau
_d=\frac{\phi _{cl}}{\dot{\phi}_{cl}}$. The
Hubble timescale is given by: 
\begin{equation}
\nu \equiv \frac{\tau _d}{\tau _H}=\tau _dH=\frac{H\phi _{cl}}{\dot{\phi}%
_{cl}}=\sqrt{\frac 23}\frac{2\pi }{M_p}\frac{\phi _{cl}}{\dot{\phi}_{cl}}%
\rho ^{1/2}.
\end{equation}
The number of e-folds in a given period of time is given by: 
\begin{equation}
N_c=\int_{t_0}^{t_0+\delta t}dt\;H=\int_{\phi _0}^{\phi _{cl}}d\phi
_{cl}^{\prime }\;\frac \nu {\phi _{cl}^{\prime }}.
\end{equation}

When the scalar field potential is sufficiently flat the field
rolls to the minimum of the potential very slowly, and the following
conditions are fulfilled
\begin{eqnarray}
\Theta &=& \frac{2}{K^2} \left( \frac{H^{\prime}}{H} \right)^2 \ll 1, \\
\Sigma &=& \frac{2}{K^2} \frac{H^{\prime\prime}}{H} \ll 1,
\end{eqnarray}
with $K={\frac{\sqrt{8 \pi} }{M_p}}$\cite{copeland}. 
When the slow-roll conditions hold, the eq. (\ref{r6}) can be
approximated by a first-order equation of motion for $\phi$\cite{Mez}.
However, in this paper we will consider the exact treatment for the 
dynamics of $\phi$.
At the end of
inflation, when the scale factor stops accelerating, one obtains
$\Theta(\phi_{cl})=1$, which determines $\phi^{end}_{cl}$.

Hence, we have $\dot{\phi}_{cl}\simeq -\frac{V^{\prime }(\phi _{cl})}{3H}$
and $H^2=\frac{8\pi ^2}{3M_p^2}V(\phi _{cl})$, so that: 
\begin{equation}
\nu \sim -\frac{8\pi ^2\phi _{cl}}{M_p^2}\frac{V(\phi _{cl})}{V^{\prime
}(\phi _{cl})},
\end{equation}
and 
\begin{equation}
N_c=-\frac{8\pi ^2}{M_p^2}\int_{\phi _0}^{\phi _{cl}}d\phi _{cl}^{\prime }\;%
\frac{V(\phi _{cl}^{\prime })}{V^{\prime }(\phi _{cl}^{\prime })}.
\end{equation}
The solution to the horizon problem requires $N_c\gtrsim 60$, which
in general implies that $\tau _d>\tau _H$.
When that scale crosses the
Hubble radius,
each length scale $\Omega _c$ is
associated with a unique value of $\phi _{cl}$
denoted by $\Omega (\phi _{cl})$.
This is given by\cite{copeland}: 
\begin{equation}
\Omega _c(\phi _{cl})=
\frac{\exp {[N_c(\phi _{cl})]}}{H(\phi _{cl})}\frac{a_o}{a_e},
\end{equation}
with $a(\phi _{cl})=a_e\exp {[-N_c(\phi _{cl})]}$.

The study of the quantum component is simplified
if we redefine the field $\phi $ with the map
$\phi =e^{-\frac 32\int dt\ H}\chi $. The equation of motion for
the field operator $\chi $ is: 
\begin{equation}
\ddot{\chi}-\frac 1{a^2}\nabla ^2\chi -\frac{k_0^2}{a^2}\chi =0 ,  \label{b}
\end{equation}
where $k_0^2=a^2\left( \frac 94H^2+\frac 32\dot{H}-V''_c\right) $.
Thus $\chi $ can be interpreted as a free scalar field with
a time dependent mass parameter.
The field $\chi$
can be expanded in a set of modes $\xi _k(t)e^{i\vec{k}.\vec{r}}$: 
\begin{equation}\label{chi}
\chi (\vec{r},t)=\frac 1{(2\pi )^{3/2}}\int
d^3k\left[ a_k\xi _k(t)e^{i\vec{k}.\vec{r}}+h.c.\right],
\end{equation}
where the annihilation and creation operators satisfy the usual commutation
relations for bosons: 
\begin{eqnarray}
&& \lbrack a_k,a_{k^{\prime }}^{\dagger }]=\delta^{(3)}
(\vec{k}-\vec{k}^{\prime}) \\
&& [a_k,a_{k^{\prime }}]=[a_k^{\dagger },a_{k^{\prime
}}^{\dagger }]=0,
\end{eqnarray}
and the modes are defined for the equation of motion 
\begin{equation}
\ddot{\xi}_k+\omega _k^2\xi _k=0,  \label{modes}
\end{equation}
with $\omega _k^2=a^{-2}\left( k^2-k_0^2\right) $. The function $k_0^2(t)$
gives the threshold between an unstable infrared sector ($k^2\ll k_0^2$), which
includes only wavelengths longer than the Hubble radius, and a
stable short wavelength sector ($k^2 \gg k_0^2$). We adopt the normalization
condition $\xi _k\dot{\xi}_k^{*}-\xi^*_k\dot\xi _k=i$ for the modes,
such that the field operators $\chi $ and $\dot{\chi}$ satisfy the
canonical commutation relations.

The coarse-grained field that describes the super Hubble spectrum contains
only the modes with wavenumber smaller than the Hubble's wavenumber $k_0$.
This field can be written as a Fourier expansion in terms of the
modes
\begin{equation}\label{cg}
\chi_{cg} (\vec{r},t)=\frac 1{(2\pi )^{3/2}}\int
d^3k \  \theta(k-\epsilon k_0) \  \left[ 
a_k\xi _k(t)e^{i\vec{k}.\vec{r}}+h.c.\right],
\end{equation}
where $\theta$ denotes the Heaviside function and $\epsilon \ll 1$ is
a dimensionless constant.

The eq. (\ref{chi}) for $\chi_{cg}$
can be written as
\begin{equation}\label{alpha}
\ddot\chi_{cg} - \mu^2(t) \chi_{cg} + \xi_c(\vec x,t)=0,
\end{equation}
where $\xi_c(\vec x,t) = -\epsilon \left[{d\over dt}\left(\dot k_0 \eta\right)
+2\dot k_0 \  \kappa\right]$, with
\begin{eqnarray}
\eta &=&\frac{1}{(2\pi)^\frac{3}{2}}\int d^3k\ \delta(\epsilon k_0 - k)
\left[ a_k e^{i\vec k.\vec r} \xi_k(t)+h.c.\right] \ , \label{eta} \\
\kappa &=&\frac{1}{(2\pi)^\frac{3}{2}}\int d^3k\ \delta(\epsilon k_0 - k)
\left[a_k e^{i\vec k.\vec r}\dot\xi_k(t)+h.c.\right]. \label{kappa}
\end{eqnarray}
These noise arises from inflow of short-wavelength modes, produced
by the cosmological evolution of both,
the horizon and the scale factor of the universe. 
Note that in eq. (\ref{alpha}) we have neglected the 
term with $-{1\over a^2} \nabla^2 \chi_{cg}$ because in the
infrared sector the following constraint is fulfilled:
$k^2/a^2 \ll k^2_0/a^2$.

\section{Wave function for the coarse-grained field}

The effective Hamiltonian related to eq. (\ref{alpha})
is
\begin{equation}\label{H}
H_{eff}(\chi_{cg},t) = \frac{1}{2} P^2_{cg}- \frac{\mu^2}{2} \chi^2_{cg}
+ \xi_c \chi_{cg},
\end{equation}
where $P_{cg} = \dot\chi_{cg}$ and $\mu^2(t)=k^2_0/a^2$. Hence, we can
write the following Schr\"odinger equation
\begin{equation}\label{gamma}
i\frac{\partial}{\partial t} \Psi(\chi_{cg},t) = - \frac{1}{2}
\frac{\partial^2}{\partial \chi^2_{cg}} \Psi(\chi_{cg},t) +
\left[-\frac{\mu^2}{2} \chi^2_{cg} + \xi_c \chi_{cg}\right] \Psi(\chi_{cg},t),
\end{equation}
where $\Psi(\chi_{cg},t)$ is the wave function of the system.
The probability density
to find the universe with the configuration $(\chi_{cg},t)$,
is
\begin{equation}
P(\chi_{cg},t) = \Psi(\chi_{cg},t) \Psi^*(\chi_{cg},t),
\end{equation}
where the asterisk denotes the complex conjugate. An elegant way to solve
the eq. (\ref{gamma}) is based on the use of explicitely time dependent
invariants of motion for time dependent quadratic potentials. Such
quantities appeared in accelerator theory\cite{36} but were first
analyzed by Lewis\cite{37}.
An invariant of the form
\begin{equation}
I(t) = A(t) \chi_{cg} + B(t) P_{cg} + C(t),
\end{equation}
is proposed, where 
the time dependent coefficients $A$, $B$, $C$ are determined by the condition
\begin{equation}
\frac{\partial}{\partial t} I(t) - i \left[I(t), H_{eff}\right]=0.
\end{equation}
This requires
\begin{eqnarray}
&& \dot A + \mu^2(t) B =0,\label{a1}\\
&& A+\dot B =0,\label{dos}\\
&& \dot C - B \xi_c =0.
\end{eqnarray}
The equations (\ref{a1}) and (\ref{dos})
can be combined to give (for $\mu^2 = {k^2_0 \over a^2}$)
\begin{equation}\label{uno}
\ddot B - \mu^2 B =0,
\end{equation}
and, since
\begin{equation}\label{iuju}
C(t) = \int^{t}_{} dt' \  B(t') \  \xi_c(\vec x,t'),
\end{equation}
the only equation that we need to solve is (\ref{uno}). From eq. (\ref{dos}),
the invariant now can be written as
\begin{equation}
I(t) = B(t) P_{cg} - \dot B \chi_{cg} + C(t).
\end{equation}
In order for $I(t)$ to be an Hermitian operator, only real solutions
of eq. (\ref{uno}) are admissible. The eigenvalue equation for this
operator is
\begin{equation}
I(t) \Phi_{\lambda} (t) = \lambda \Phi_{\lambda}(t),
\end{equation}
where the eigenfunction\cite{39}
\begin{equation}
\Phi_{\lambda}(t) = \frac{1}{\sqrt{2\pi B}} \  e^{\frac{i}{B}
\left[\frac{1}{2}\dot B \chi^2_{cg} + (\lambda - C)\chi_{cg}\right]},
\end{equation}
satisfies the orthogonality condition
\begin{equation}
\left<\Phi_{\lambda}(t) \left| \Phi_{\lambda '}(t) \right.\right>=
\delta(\lambda - \lambda ').
\end{equation}
Here, the eigenvalues $\lambda$ are independent of the time. Since the
action of $I(t)$ does not involve time derivatives, the function
$\Phi_{\lambda}(t)$ is always arbitrary up to a time dependent
phase factor. Hence, we can write the eigenfuctions of $I(t)$
\begin{equation}\label{ddos}
\psi_{\lambda}(t) = e^{i\alpha_{\lambda}(t)} \Phi_{\lambda}(t).
\end{equation}
Taking the time derivative in eq. (\ref{ddos}),
is obtained
\begin{equation}\label{tres}
\frac{\partial I}{\partial t} \psi_{\lambda}(t) =
(\lambda - I) \frac{\partial}{\partial t} \psi_{\lambda}(t).
\end{equation}
This can be written in terms of $\Phi_{\lambda}(t)$
as
\begin{equation}
\left<\Phi_{\lambda}(t) \left| \left(H_{eff} - i \frac{\partial}{\partial t}
\right) \right| \Phi_{\lambda}(t) \right> =
-\frac{\partial}{\partial t} \alpha_{\lambda}(t) \left<
\Phi_{\lambda}\left| \Phi_{\lambda}\right.\right>,
\end{equation}
where $\alpha_{\lambda}(t)$ comes from solving
\begin{equation}
\alpha_{\lambda}(t) = - \Large{\int}^t dt' \frac{[\lambda - C(t')]^2}{
2 B^2(t')}.
\end{equation}
The solutions of the time dependent Schr\"odinger equation are
\begin{equation}
\psi_{\lambda}(t) = \frac{1}{\sqrt{2\pi B}} e^{\frac{i}{B}
\left[\frac{1}{2} \dot B \chi^2_{cg} + (\lambda - C)\chi_{cg} + \beta_{\lambda}
\right]},
\end{equation}
where
\begin{equation}
\beta_{\lambda}(t) = B(t) \alpha_{\lambda}(t).
\end{equation}
The general solution $\Psi(\chi_{cg},t)$ can be represented as
\begin{equation}
\Psi(\chi_{cg},t) = \Large{\int} d\eta \  \eta_{\lambda}
\psi_{\lambda}(\chi_{cg},t),
\end{equation}
where $\eta_{\lambda}$ the coefficients are
\begin{equation}
\eta_{\lambda} = \left< \psi_{\lambda}(t_0)\left| \Psi(t_0)
\right.\right>,
\end{equation}
and $t_0$ is the initial time. In our case $t_0$ is the time when
inflation starts, and corresponds to $1$ in Planckian unities.
We will take $\beta_{\lambda}(t_0) = 0$, $B(t_0)=1$, $\dot B(t_0)=0$,
and $C(t_0)=0$. With these choice is obtained
\begin{equation}
\eta_{\lambda} = \frac{1}{\sqrt{2\pi}} {\Large{\int}}
d\chi_{cg} e^{-i\lambda \chi_{cg}} \Psi(\chi_{cg},t_0),
\end{equation}
which is the Fourier transform of the initial state. In the case
of an squeezed harmonic oscillator, $\Psi(\chi_{cg},t_0)$ can
be written as
\begin{equation}
\Psi_{s}(\chi_{cg},t_0) = \left[\frac{1}{2\pi \sigma^2}\right]^{1/4}
e^{-\left[\frac{\chi_{cg}-\chi^{(0)}_{cg}}{2\sigma}\right]^2 +
i P^{(0)}_{cg} \chi_{cg}}.
\end{equation}
Here, $\sigma$ is the width of the gaussian and $P^{(0)}_{cg} = \dot
\chi_{cg}(\vec x,t_0)$.
By setting $z=P^{(0)}_{cg} - \lambda$, the wave function becomes
\begin{eqnarray}
\Psi(\chi_{cg},t) &=& \left[\frac{\sigma}{\sqrt{2\pi} \pi B}\right]^{1/2}
e^{\frac{i}{B}\left[\frac{1}{2} \dot B \chi^2_{cg}
+ (P^{(0)}_{cg}-C)\chi_{cg}\right]} \\
&\times & \Large{\int} dz \  e^{-\sigma^2 z^2 + i(\chi^{(0)}_{cg}-
\chi_{cg}/B)z+i\beta_z/B},
\end{eqnarray}
where $\beta_z$ is given by
\begin{equation}
\beta_z = B(t) \left[-z^2 {\cal R}(t) + z {\cal U}(t) - {\cal S}(t)\right],
\end{equation}
being
\begin{eqnarray}
{\cal R}(t) & = & \Large{\int}^t dt' \frac{1}{2 B^2(t')}, \\
{\cal U}(t) & = &\Large{\int}^t dt' \frac{[P^{(0)}_{cg} - C(t')]}{
B^2(t')},\\
{\cal S}(t) & = & \Large{\int}^t dt' \frac{[P^{(0)}_{cg} - C(t')]^2}{
2 B^2(t')}.
\end{eqnarray}
In our case, $\chi_{cl} = \left<\chi_{cg}\right>$ and $P_{cl} =
\left<\dot\chi_{cg}\right>$ imply that
\begin{eqnarray}
\chi_{cg}(t) &=& B(t) \left[ \chi^{(0)} + {\cal U}(t)\right],\\
P_{cl} & = & \frac{1}{B} \left[\dot B \chi_{cl} + P^{(0)}_{cg} - C(t) \right].
\end{eqnarray}
Furthermore, we can define the parameter
\begin{equation}
\Delta^2(t) = \frac{B^2(t)}{\sigma^2} \left[\sigma^4 +
{\cal R}^2(t)\right],
\end{equation}
such that the wave function can be written as
\begin{eqnarray}
\Psi(\chi_{cg},t) & = & \frac{1}{(2\pi)^{1/4} \Delta^{1/2}}
e^{-\frac{1}{4 \Delta^2}\left[\chi_{cg} - \chi_{cl}\right]^2}
e^{i\frac{\chi^2_{cg}}{\Delta^2} \left[2 \frac{\dot B}{B} \Delta^2
+ \frac{{\cal R}(t)}{\sigma^2}\right]} \\
&\times & e^{i \frac{\chi_{cg}}{\Delta^2} \left[\Delta^2\left(P_{cl} -
\frac{\dot B}{B} \chi_{cl}\right) - \frac{{\cal R}(t) \chi_{cl}}{2 \sigma^2}
\right]} e^{i \gamma(t)},
\end{eqnarray}
where $\gamma(t)$ is an arbitrary phase. Furthermore, it
can be verified
that the expectation value of the effective energy in the infrared sector is
\begin{equation}\label{cuatro}
\left<E_{eff}\right> = E_{cl} + \frac{1}{8\Delta^2} -
\frac{\mu^2}{2} \Delta^2 + \frac{1}{2} \left( \frac{\dot B}{B}
\Delta + \frac{{\cal R}}{2\Delta \sigma^2}\right)^2,
\end{equation}
where
\begin{equation}\label{E}
E_{cl} = \frac{1}{2} P^2_{cl} - \frac{\mu^2}{2} \chi^2_{cl} 
+ \xi_c \  \chi_{cl}.
\end{equation}
The contribution of the quantum fluctuations to $\left<E_{eff}\right>$ 
is represented
by the second, third, and fourth terms of (\ref{cuatro}).
The fourth term describes the decoherence of the system, in such away 
that it
is zero for coherent states and positive for 
decoherentized states. The parameter $\mu=k_0/a$ depends on the
cosmological model. On the other hand, the squared fluctuations 
$\left<\chi^2_{cg}\right>$ and $\left<P^2_{cg}\right>$ are
\begin{eqnarray}
\left<\chi^2_{cg}\right> & = & \chi^2_{cl}(t)+ \Delta^2(t), \label{al}\\
\left<P^2_{cg}\right> & = & P^2_{cl}(t) + \frac{1}{4 \Delta^2(t)} +
\left(\frac{\dot B}{B} \Delta(t) + \frac{{\cal R}}{2\Delta \sigma^2}
\right)^2.\label{be}
\end{eqnarray}

\section{Wave function in a power-law expansion of the universe}

If mass parameter $\mu$ is time independent but the external
classical force is nonzero, the solutions are coherent states\cite{25}.
This case corresponds to a de Sitter expansion of the universe, which
was studied in a previous paper by Habib and Miji\'c\cite{HM}.
In this work we are interested in the study of the
particular case of a power-law expansion of the universe in which the scale
factor is
$a \propto t^p$. In this case the squared parameter of mass is given 
by\cite{BCMS}
\begin{equation}
\mu^2(t) = M^2 \  t^{-2},
\end{equation}
where $M^2 ={9\over 4} p^2 -{15\over 2}p +2 $
and the Hubble parameter being given by $H(t)=p/t$\cite{BCMS}.
The condition to get an
unstable sector is $M^2>0$, or $p> (5+\sqrt{17})/3 \simeq 3.04$.
Furthermore, since $H(\phi_c) = H_0 \  e^{\phi/M_p}$, the slow-roll
paramters $\Theta$ and $\Sigma$ become both of the order of
$10^{-1}$, so that the
slow-roll regime is guaranted.
Hence, the
equations which characterize the system are
\begin{eqnarray}
&& \ddot B - M^2 t^{-2} B =0, \label{A}\\
&& C(t) = \Large{\int}^t dt' \  B(t') \xi_c(\vec x,t'). \label{B}
\end{eqnarray}
The general solution of eq. (\ref{A}) is
\begin{equation}
B(t) = c_1 t^{\frac{1}{2}[1+ \sqrt{1+4M^2}]} +
c_2 t^{\frac{1}{2}[1- \sqrt{1+4M^2}]},
\end{equation}
where the initial conditions $B(t_0)=1$ and
$\dot B(t_0)=0$ imply
\begin{equation}
c_1 = \frac{\sqrt{1+4M^2}-1}{2\sqrt{1+4M^2}}, \qquad
c_2=  \frac{\sqrt{1+4M^2}+1}{2\sqrt{1+4M^2}}.
\end{equation}
Since it is difficult to know
exactly the functions ${\cal R}(t)$, ${\cal U}(t)$
and ${\cal S}(t)$, we can make the calculation for late
times. For $t\gg 1$, one obtains $\left.B(t)\right|_{t\gg 1}
\simeq t^{1/2}/2$ and 
\begin{eqnarray}
&& \left.{\cal R}(t)\right|_{t\gg 1} \simeq 
\frac{\sqrt{1+4M^2}}{(\sqrt{1+4M^2}-1)(1+\sqrt{1+4M^2})} \  {\rm ln}
\left[\frac{2 \sqrt{1+4M^2}}{\sqrt{1+4M^2}-1}\right], \label{R}\\
&& \left.\Delta^2(t)\right|_{t\gg 1} \propto t^{1+\sqrt{1+4M^2}}.
\label{be1}
\end{eqnarray}
This means that the decoherence function ${\cal D}(t) =
{1\over 2} \left({\dot B \over B} \Delta +
{{\cal R}\over 2 \Delta \sigma^2}\right)^2$ in the expectation effective
energy (\ref{cuatro}), will be
\begin{equation}\label{al1}
\left.{\cal D}(t)\right|_{t\gg 1}
\propto t^{\frac{1}{2} \left(\sqrt{1+4M^2}-1\right)},
\end{equation}
which always increases because $M^2 >0$ during inflation. 
Replacing (\ref{R}), (\ref{be1}) and (\ref{al1})
in eqs. (\ref{al}) and (\ref{be}) one observes that, 
since $\left.\Delta^2(t)\right|_{t\gg 1}$ increases with time, $\left<
\chi^2_{cg}\right>$ increases but the second term in (\ref{be}) decreases
as $t$ increases. The second terms in (\ref{al}) and (\ref{be})
describe the evolution of super Hubble fluctuations due to the 
exchange of degrees of freedom between the infrared ($k^2 \ll k^2_0$)
and the ultraviolet ($k^2 \gg k^2_0$) sectors. The third term in
eq. (\ref{be}) is the most interesting one. This term describes
the decoherence of the
phase-space ($\chi_{cg},P_{cg}$) during inflation. Note
[see eq. (\ref{al1})], that it increases with $t$, 
so the phase-space
loses its coherence at the end of inflation. This term comes from the
coupling between $\chi_{cg}$ and $P_{cg}$. 
This is the physical origin
of decoherence in the phase-space ($\chi_{cg},P_{cg}$). 
However, the relevant phase-space during inflation is 
($\phi_{cg},P_{\phi_{cg}}$), where $\phi_{cg} = a^{-3/2} \chi_{cg}$ and
$P_{\phi_{cg}}=\dot\phi_{cg}$. This implies that the function that
describes decoherence between $\phi_{cg}$ and $P_{\phi_{cg}}$ can be
written, to a good approximation, as
\begin{equation}
\left.{\cal D}\left(\phi_{cg},P_{\phi_{cg}},t\right)\right|_{t\gg 1} \sim
\left. a^{-3} {\cal D}(t)\right|_{t\gg 1} \propto t^{\frac{1}{2}\left[
\sqrt{1+4M^2}-\left(6p+1\right)\right]},
\end{equation}
where $\left.{\cal D}(t)\right|_{t\gg 1}$ is given by (\ref{al1}).
A numerical calculation shows that ${\cal D}(\phi_{cg},P_{\phi_{cg}},t)$ grows
for $p > 4.6$, but decreases in the range $3.04 < p < 4.6$. It is
well known that power-law inflation takes place if $p>3.04$. This means that
for $3.04 < p < 4.6$ the
phase-space ($\phi_{cg},P_{\phi_{cg}}$) remains coherent during inflation.

\section{Final Remarks}

In this paper we have considered 
the wave function for the coarse-grained field. It
describes the redefined matter field fluctuations in the infrared 
sector for a globally flat FRW background metric.
The Hamiltonian of the Schr\"odinger equation is given by an effective
Hamiltonian. It has a quadratic contribution with a time dependent 
mass plus an effective stochastic force, which describes the ``interaction''
between both, ultraviolet and infrared sectors. However, this
is not a true interaction in the sense of a thermalyzed
environment. Here, continuosly we have new modes are crossing the
ultraviolet sector increasing the number of degrees of freedom of 
the infrared
sector. This effect appears in the stochastic equations as an effective
noise $\xi_c$, which is responsible for the uncertaintly of the
quantum state.

On the other hand the lost of coherence in the phase-space is a 
consequence of a nonzero correlation between the variables of such
a space (in this case $\chi_{cg}$ and $P_{cg}$). As was shown in this
work, the decoherence function ${\cal D}(t)$ increases with time
in such away that the phase-space ($\chi_{cg},P_{cg}$) is
decohered at the end of power-law inflation.
However, the relevant phase-space to describe decoherence during inflation
is ($\phi_{cg},P_{\phi_{cg}}$). We found that
in the range $3.04 < p < 4.6$ the system does not decohere.
This implies that, under these conditions, the classical treatment of
super Hubble matter field fluctuations developed in stochastic inflation
should be revised.
On the other hand for very large $p$ (more exactly for $p > 4.6$) the
phase-space ($\phi_{cg},P_{\phi_{cg}}$) loses its coherence at the end
of inflation, and then stochastic inflation provides a good treatment
for large-scale matter field fluctuations when the scale factor grows
very rapidly.
This
calculation was performed in the framework of supercooled inflation, which
does not take into account dissipative effects produced by the interaction
between the
inflaton field and other particles of a thermal bath. 
In the
warm inflation\cite{Berera,Bellini}, 
the phase-space could not either remain coherent, but as a consequence
of additional thermal and dissipative effects. 
A more detailed treatment has to deal
with complicated nonlinear effects of super Hubble fluctuations.
This goes 
beyond the scope of this
paper. I hope to consider this topic elsewhere.\\

\centerline{\bf ACKNOWLEDGMENTS}
\noindent
I would like to acnowledge F. Astorga for his careful reading of the 
manuscript.
Also, I would like to acknowledge 
CONACYT (M\'exico) and CIC of Universidad Michoacana
for financial support in the form of a research grant.\\

\end{document}